# Compress Voice Transference over low Signal Strength in Satellite Communication

Saira Beg [1], M. Fahad Khan [1], Faisal Baig [1]
[1]COMSATS Institute of Information and Technology
[2] Federal Urdu University of Arts, Science and Technology,
Islamabad, Pakistan.
sairabegbs, mfahad.bs,engr.fsl.baig@gmail.com,

**Abstract**

This paper presents the comparison of compression algorithms for voice transferring method over SMS in satellite communication. Voice transferring method over SMS is useful in situations when signal strength is low and due to poor signal strength voice call connection is not possible to initiate or signal dropped during voice call. This method has one serious flaw that it produces large number of SMS while converting voice into SMS. Such issue is catered to some extend by employing any compression algorithm. In this paper our major aim is to find best compression scheme for said method, for that purpose we compare 6 different types of compression algorithms which are; LZW (Lempel-Ziv-Welch), Huffman coding, PPM (Prediction by partial matching), Arithmetic Coding (AC), BWT (Burrows-Wheeler-Transform), LZMA (Lempel-Ziv-Markov chain).

This comparison shows that PPM compression method offers better compression ratio and produce small number of SMS. For experimentation we use Thuraya SG-2520 satellite phone. Moreover, we develop an application using J2ME platform. We tested that application more than 100 times and then we compare the result in terms of compression ratio of each algorithm and number of connected SMS produce after each compression method. The result of this study will help developers to choose better compression scheme for their respective applications.

1. Introduction

Mobile communication becomes the essential part for today's daily life. Such systems revolutionized the way of interaction and have ability to send data from almost arbitrary locations. The most famous and widely deployed standard of mobile communication is GSM (Global System for Mobile communications). GSM gain lot of attention and have 4.4 billion subscribers around the world [1]. To extend such mobility in under-populated regions, in war, ships, on planes, mountains where no terrestrial/GSM communications infrastructure is available there satellite communication is the only viable solution. Such communication carried through satellite orbiting around the earth rather than towers. Satellite transmission depends upon atmospheric transport. Climate changes (rain and cloudy), tall buildings and dense forest etc. can affect the transmission by losing signal strength. With weak signal strength making a voice call is difficult so sending SMS could be the best option here [2-3].

Voice transferring method over SMS is an alternative way of sending audio messages via satellite communication system. This method converts the audio messages into characters and then sets these characters as a payload text of SMS. One serious weakness of above method is that it produces large number of SMS [4]. Such issue is catered to some extend by employing compression algorithm. In literature several types of lossless compression algorithms are available; selection of best compression scheme for said method is depend upon the two terms; compression ratio and number of SMS. This paper provides a brief comparison of six different lossless compression algorithms for the said method. This Paper is divided into following sections. Section 2, 3 presents the brief detail of compression algorithms and methodology respectively, section 4 represents performance metrics and dataset, section 5 is about results discussion and comparison, and at the last we conclude the paper.

2. Compression Algorithms

2.1. LZW (Lempel-Ziv-Welch)

http://www.learnrnd.com/news.php?id=ISSUES_IN_MOLECULAR_COMMUNICATIONS

Lempel-Ziv-Welch (LZW) algorithm was proposed by Lempel Jakob Ziv and then modified by the Terry Welch in 1984. Generally LZ family algorithms have used the dictionary based method in which dictionary of previous encoded strings are stored [5]. Dictionary indexes could be of various sizes but most useful are; 512 and 16384, which uses 9 and 14 bits respectively for the index representation. Moreover the first 256 index entries are usually reserved for symbols [5-6].

LZW algorithm starts with scanning the whole file and search for such data sequences that occurs more than once. Now stored these sequences into the dictionary and put references (symbols) within the compressed file where ever repetitive sequence occurred. LZW compression scheme is an adaptive approach because changing dictionary of the strings that have appeared in the text so far is maintained and it guaranteed that every input sequence (input file) should be converted into the dictionary indexes (symbols) [6].

## 2.2. LZMA (Lempel-Ziv-Markov Chain)

LZMA algorithm is the one of the refined version of LZ family which combines the fundamentals of LZ77 compression scheme along with the Markov Chain and a variant of Arithmetic Coding called range encoder. It first used for 7-zip compression software in 2001[7-9]. LZMA used the literals and phrases instead of byte-based structure which helps to avoid the mixing of unrelated contents. Because of this property it gains better compression ratio over pervious LZ algorithms [8]. Moreover it has various dictionary sizes range up to 4GB having fast compression/decompression rate [9].

## 2.3. Huffman Coding (HC)

Huffman Coding algorithm was developed by David Huffman in 1950. Huffman coding algorithm contains a set of probabilities which used to generate prefix codes. Following steps are used to generate prefix-code tree;

- Algorithm starts with the forest of trees. Each tree belongs to single a message having single vertex with the weight of $w_i = p_i$
- Repeat the following tasks until a single tree remains;

- Select two tree having lowest weighted roots ($w_1$ and $w_2$)
- Join the both trees by adding a larger weighted root ($w_1+w_2$) and make the tree its Childs Code-words are assigned according to probability values. Code-words are small if the probability is high and longer code-words used when probability is low. Huffman coding algorithm is the simplest compression algorithm which used as back end of JPEG, GZIP and for many other standards as well [10-11].

### 2.4. PPM (Prediction by Partial Matching)

A PPM compression algorithm belongs to the category of such algorithms which produce best compression ratios then others but generally it is not considered as fast [10]. PPM compression method takes the advantage of previous k characters to generate the conditional probability for the current character. This could be possible if we keep track (dictionary) of every strings and a count of each character occurrence in that string. Now calculates the conditional probabilities and pass it to Huffman or other coder to generate a bit sequence [10].

### 2.5. Arithmetic Coding (AC)

In Arithmetic Coding, fractions are used to represents the entire source message instead of code-words represented a symbol of the text. In this method occurrence probabilities and the cumulative probabilities of the source message is used. Cumulative probability range is used for encoding/decoding process and its range is created in the beginning. Now start reading the source file character by character and corresponding range of character is set within the cumulative probability. Then divide the range into sub parts according to the probabilities of the alphabet, now read next character and select the corresponding sub range. The whole source file is read in a similar fashion and finally a number should be taken from the final sub range as the output of the encoding process [11].

### 2.6. BWT (Burrows-Wheeler-Transform)

BWT is finest lossless compression algorithm which utilized in BZIP. It produces good compression ratios within 10% of the best methods such as PPM but it is faster than PPM.

Unlike PPM BWT use the Block Sorting Transform which reduces the time [10]. It generally converts the data into the form which can easily compressed by Run-Length encoders and other statistical encoders with order greater than 0 [12].

### 3. Methodology

Recently an alternative way of voice transferring method is presented in [4]. In this method SMS is used as transferring medium. As it already known that SMS is limited to text only and it could not send/receive any multimedia content e.g. voice, image, sound etc. so if one can convert voice data into text than it can sent through SMS. Voice transferring method over SMS used the same idea. In that method authors converts the audio message into ByteArrayOutputStream. After that they convert ByteArrayOutputStream into unsigned integer array. From where they get Extended ASCII characters by converting unsigned integer array into their respective characters. In their method they used LZW compression scheme as well. But still major issue with this method is that it produces large number of SMS.

For choosing the best compression algorithm for such kind of applications we compare 6 different compression schemes in order to select best of them. Figure 1 shows the block diagram of proposed methodology for voice transferring method over satellite communication.

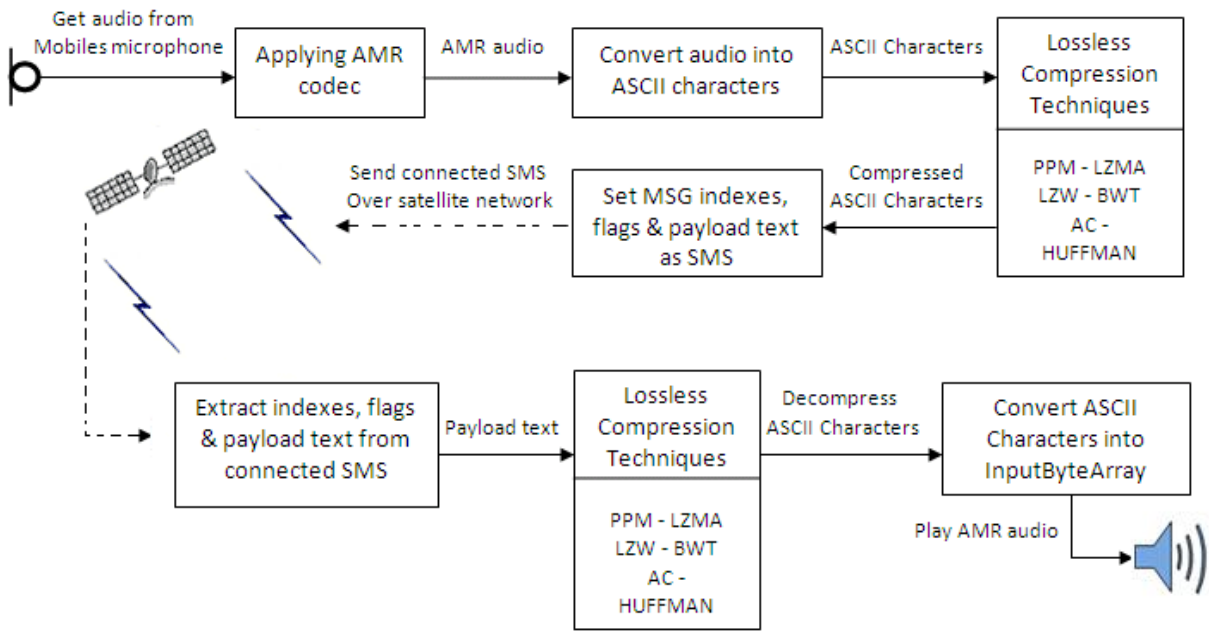

**Figure 1:** Block diagram of proposed methodology

## 4. Performance metrics and Data Set

Each compression algorithm is compared on two terms; compression ratio and number of SMS. Both terms are inversely proportional to each other. It directly means that higher the ratio rate will reduce the number of messages. Moreover compression ratio factor can also be affected by the total number of words in an audio message (length of audio message). For testing we used different length of audio messages as shown in table 1. Here, we select 9 different sentences using AMR codec mainly.

Table 1: Selected sentences & their occurrences in testing

| Sent. ID | Full Sentence | Sentence Repetition | Number of words & alphabets count |
|---|---|---|---|
| S1. | " Quick brown fox jumps over the lazy dog " | 10 times | 8, 32 |
| S2. | " Quick brown fox jumps over the lazy dog , Quick brown fox jumps over the lazy dog " | 10 times | 16, 64 |
| S3. | " Quick brown fox jumps over the lazy dog , Quick brown fox jumps over the lazy dog , Quick brown fox jumps over the lazy dog " | 10 times | 32, 96 |
| S4. | " This is a audio clip " | 10 times | 5, 16 |
| S5. | " This is a audio clip , This is a audio clip " | 10 times | 10, 32 |
| S6. | " This is a audio clip , This is a audio clip , This is a audio clip " | 10 times | 15, 48 |
| S7. | "Hello world" | 10 times | 2, 10 |
| S8. | "Hello world , Hello world" | 10 times | 4, 20 |
| S9. | "Hello world , Hello world , Hello world" | 10 times | 6, 30 |

## 5. Results and Discussions

For experiments, first we develop an application using J2ME platform for voice transferring method over satellite communication as presented in [4]. Then we apply compression methods one by one and test their performance on 9 different AMR codec sentences (see section 4). Moreover, we compare the performance on two parameters; number of characters and number of SMS which directly proportional to each other. Figure 2a presents the results of each algorithm against S1, S2 and S3 in such a way that 1st to 10th tests of the figure 'a' represent experiments having Sentence ID S1. 11th to 20th tests of the figure 'a' represent experiments having Sentence ID S2 and 21th to 30th tests of the figure 'a' represent experiments having Sentence ID S3. Similarly Figure 2b and figure 2c shows the results of S4, S5, S6, S7, S8 and S9 respectively. Results depict that AMR Codec produces larger number of characters against each sentence.

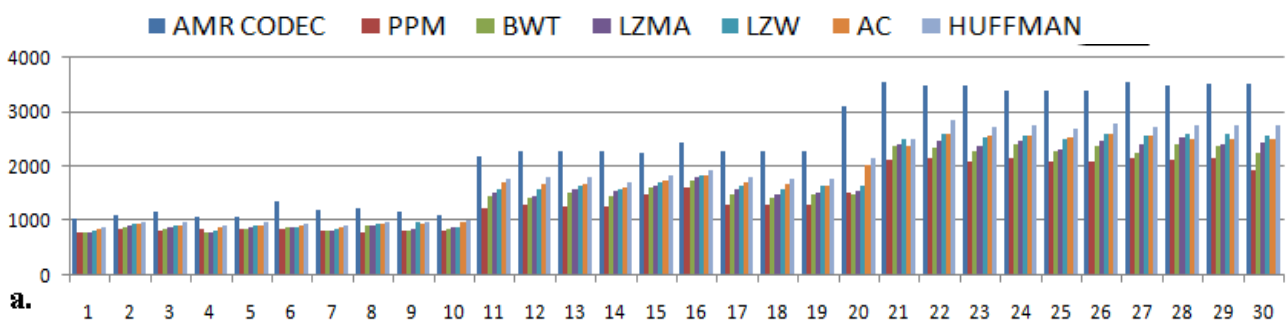

Figure 2a: number of characters against 6 compression algorithm for S1, S2 and S3

http://dx.doi.org/10.1504/IJSSE.2013.056303

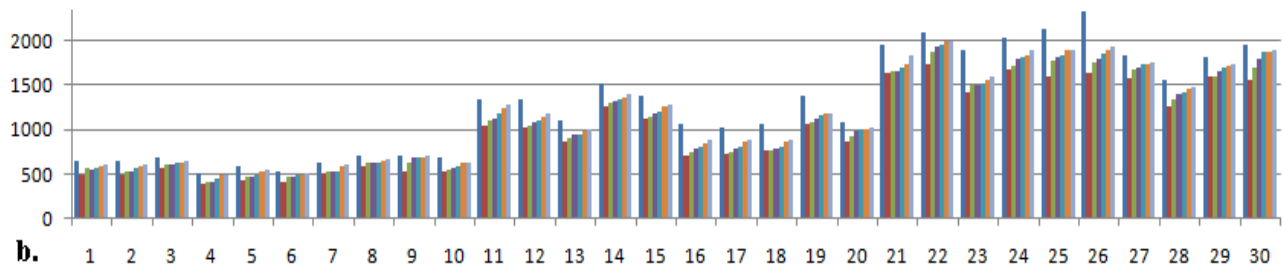

Figure 2b: number of characters against 6 compression algorithm for S4, S5 and S6

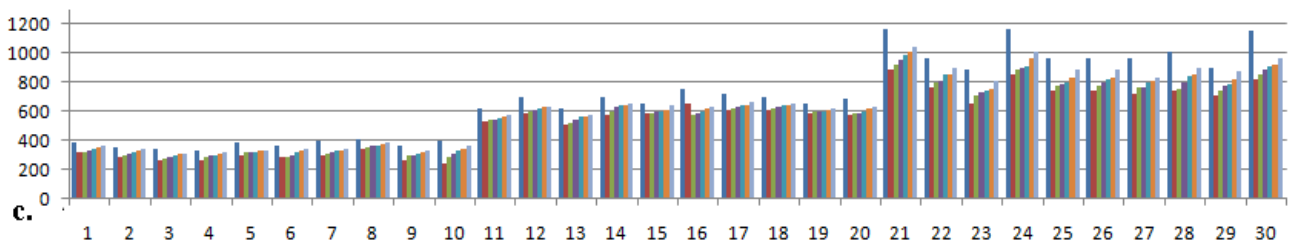

Figure 2c: number of characters against 6 compression algorithm for S7, S8 and S9

As we know that increase in number of characters produces large SMS. Figure 2d, 2e and 2f show the number of SMS against each algorithm. We clearly see that AMR Codec produces large number of SMS.

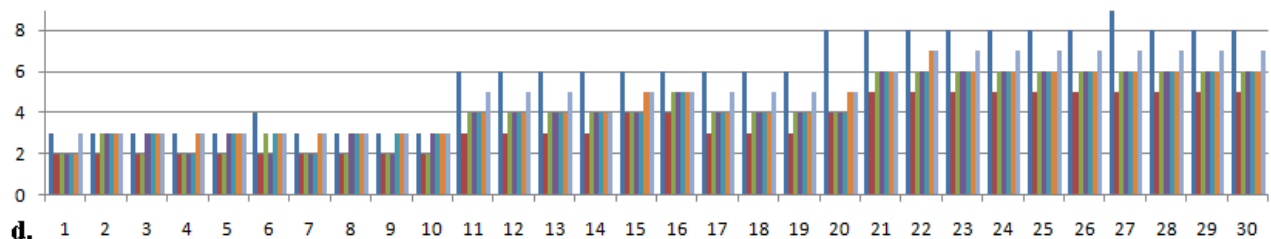

Figure 2d: number of SMS against 6 compression algorithm for S1, S2, and S3

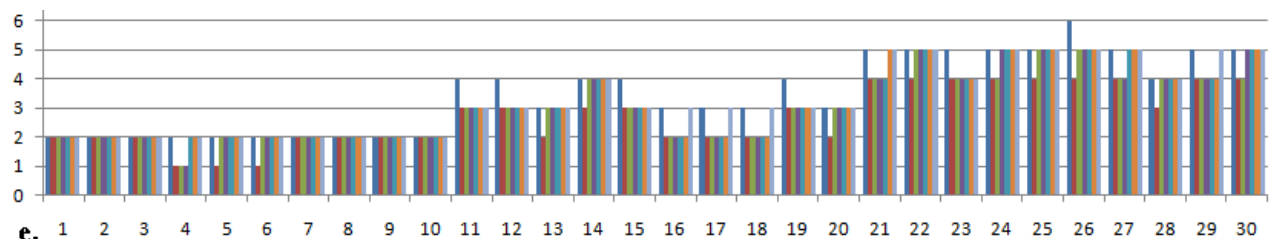

Figure 2e : number of SMS against 6 compression algorithm for S4, S5, and S6

http://www.learnrnd.com/news.php?id=ISSUES_IN_MOLECULAR_COMMUNICATIONS

http://dx.doi.org/10.1504/IJSSE.2013.056303

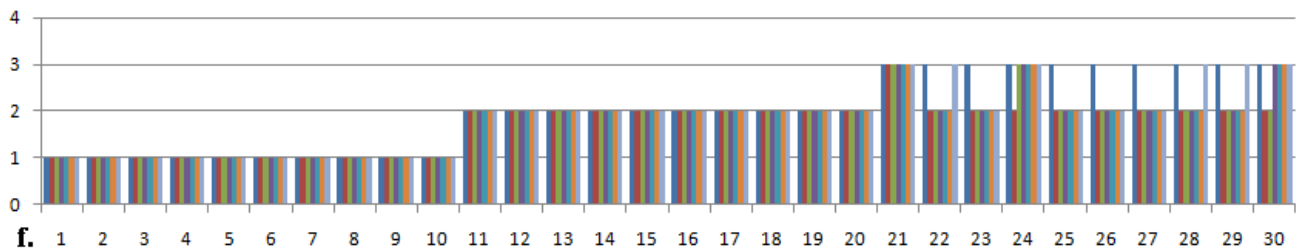

Figure 2f : number of SMS against 6 compression algorithm for S7, S8, and S9

Figure shows the comparison of each compression algorithm. We check the performance against each sentence and come to know that PPM compression algorithm performs well then other compression algorithms. So with these results we can built an hypothesis that PPM compression algorithm can produce better compression ratio if used in those applications which involves AMR codec.

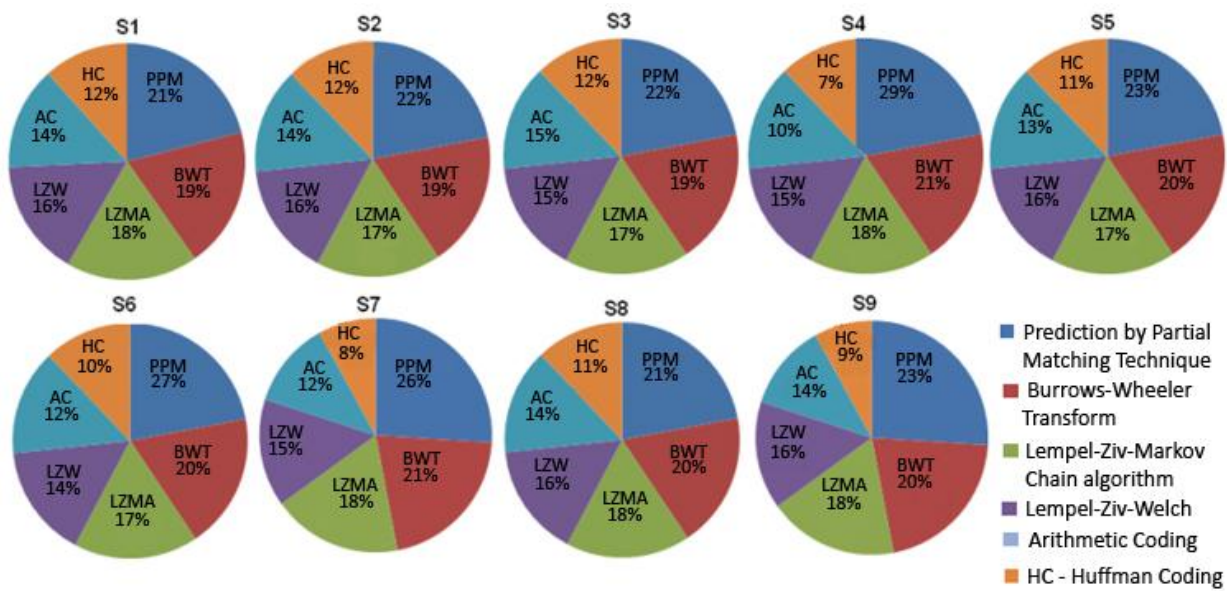

## 6. Conclusion

This paper presents the comparison of different compression algorithm for voice transferring method over satellite communication. For that purpose we compare 6 different types of compression algorithms which are; LZW (Lempel-Ziv-Welch), Huffman coding, PPM (Prediction by partial matching), Arithmetic Coding (AC), BWT (Burrows-Wheeler-Transform), LZMA (Lempel-Ziv-Markov chain). Results clearly depict that PPM compression method offers better compression ratio and produce small number of SMS then other lossless compression algorithms. The result of this study will help developers to choose better compression scheme for their respective applications.